\documentclass[10pt,twocolumn,letterpaper]{article}
\usepackage[accsupp]{axessibility}  
\usepackage{iccv}
\usepackage{times}
\usepackage{epsfig}
\usepackage{graphicx}
\usepackage{amsmath}
\usepackage{amssymb}
\usepackage{authblk}

\usepackage[pagebackref=true,breaklinks=true,letterpaper=true,colorlinks,bookmarks=false]{hyperref}

\iccvfinalcopy 

\def\iccvPaperID{19} 

\ificcvfinal\fi

\begin{document}

\title{H\&E-adversarial network: a convolutional neural network to learn stain-invariant features through Hematoxylin \& Eosin regression}

\author[1,2]{Niccol\`{o} Marini\thanks{niccolo.marini@hevs.ch}}
\author[1,3]{Manfredo Atzori}
\author[1,4]{Sebastian Ot\'{a}lora}
\author[2]{Stephane Marchand-Maillet}
\author[1,5]{Henning M\"{u}ller}
\affil[1]{Institute of Information Systems, HES-SO (University of Applied Sciences and Arts Western Switzerland)}
\affil[2]{Department of Computer Science, University of Geneva}
\affil[3]{Department of Neurosciences, University of Padua}
\affil[4]{Inselspital, Bern Hospital}
\affil[5]{Medical Faculty, University of Geneva }

\maketitle
\ificcvfinal\thispagestyle{empty}\fi

\begin{abstract}
Computational pathology is a domain that aims to develop algorithms to automatically analyze large digitized histopathology images, called whole slide images (WSI).
WSIs are produced scanning thin tissue samples that are stained to make specific structures visible. 
They show stain colour heterogeneity due to different preparation and scanning settings applied across medical centers.
Stain colour heterogeneity is a problem to train convolutional neural networks (CNN), the state-of-the-art algorithms for most computational pathology tasks, since CNNs usually underperform when tested on images including different stain variations than those within data used to train the CNN.
Despite several methods that were developed, stain colour heterogeneity is still an unsolved challenge that limits the development of CNNs that can generalize on data from several medical centers. 
This paper aims to present a novel method to train CNNs that better generalize on data including several colour variations.
The method, called H\&E-adversarial CNN, exploits H\&E matrix information to learn stain-invariant features during the training.
The method is evaluated on the classification of colon and prostate histopathology images, involving eleven heterogeneous datasets, and compared with five other techniques used to handle stain colour heterogeneity.
H\&E-adversarial CNNs show an improvement in performance compared to the other algorithms, demonstrating that it can help to better deal with stain colour heterogeneous images.
\end{abstract}
\section{Introduction}
Stain colour heterogeneity remains a critical challenge in computational pathology~\cite{TLB2019}, despite the increasing number of methods developed to tackle the problem.

\begin{figure}[t!]
\centering
\includegraphics[width=\linewidth]{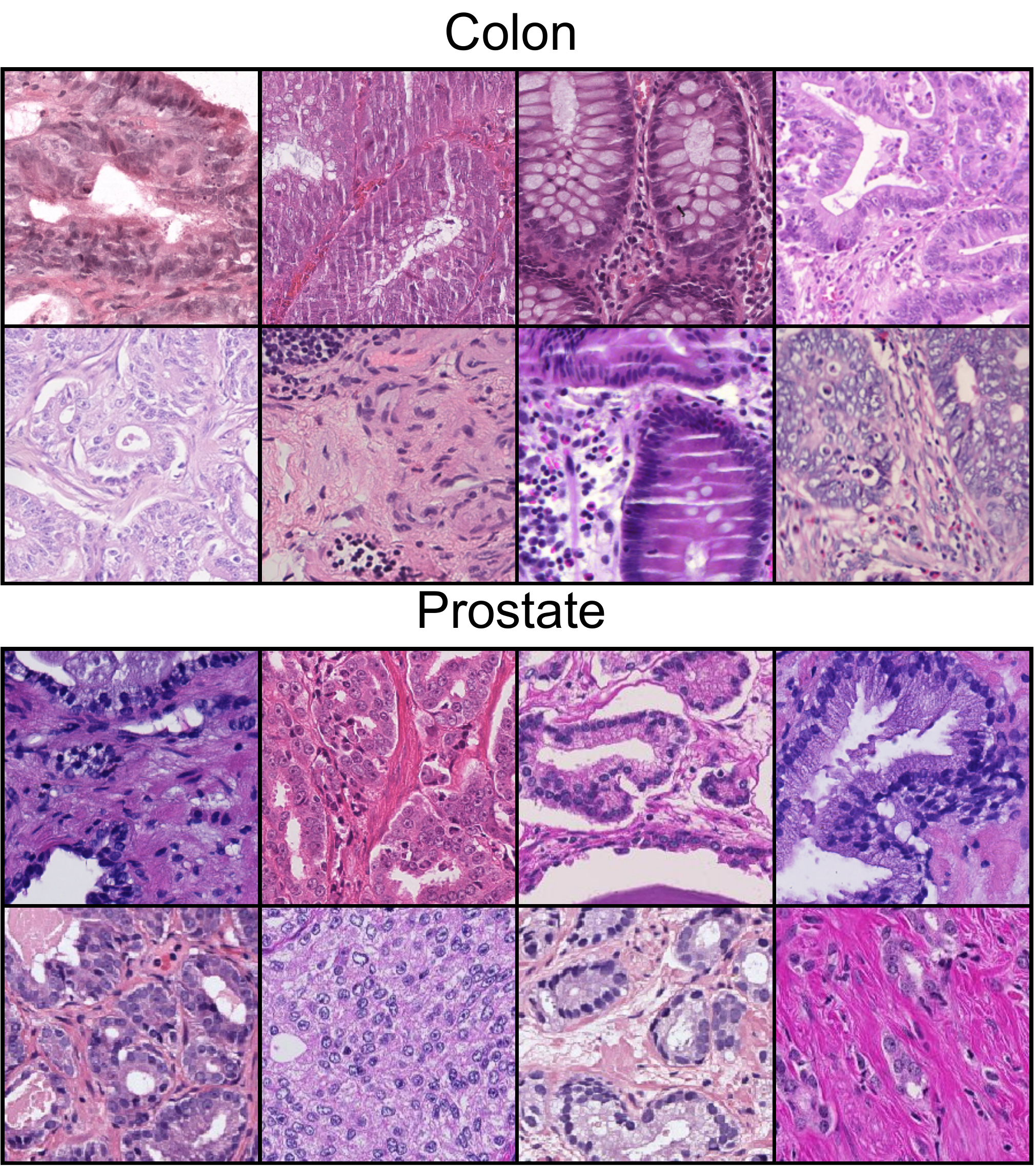}
\caption{Example of colour variations in histopathology images, collected from publicly available repositories. The figure includes examples from colon tissue (above) and prostate tissue (below).}
\label{fig:heterogeneity_images} 
\end{figure}

Stain colour heterogeneity is related to the colour variations of digitized histopathology images~\cite{VPS2016, TLB2019, LPE2019}, called whole slide images (WSI).
WSIs include slices of tissue samples collected through biopsies or surgical resections to identify findings that may lead to diseases such as cancer~\cite{GBC2009}.
Colour variations are due to the inconsistencies in sample thickness, tissue preparation and tissue scanning across different medical centers~\cite{OAA2019,SEU2020,KAO2020,RHS2019,TLB2019}.
Tissue preparation involves several procedures, including tissue staining.
The staining involves the application of chemical reagents on the sample to increase contrast among structures within the tissue.
Frequently, tissues are stained with the combination of Hematoxylin and Eosin (H\&E).
The Hematoxylin stains the cellular nuclei with shades of blue, while the Eosin stains the cytoplasm and extracellular structures with shades of pink~\cite{VPS2016}.
The concentrations of H\&E are not standardized, leading to variations of staining across medical centers.
Tissue scanning is the high-resolution digital acquisition of the image through whole slide scanners.
Several vendors produce whole slide scanners, such as Aperio, 3DHistech, Hamamatsu.
Each type of scanner presents different properties and characteristics, such as the pshysical temperature~\cite{MGK2012} (that influence the stain reagents) or the light acquired by the scanner during the acquisition, that influence the colour response of the scanner~\cite{VPS2016}.
Therefore, usually the images coming from a medical center show specific colours due to the preparation and the scanner adopted~\cite{SDB2020}.
Figure~\ref{fig:heterogeneity_images} shows examples of stain variations.

Computational pathology is a domain that aims to develop algorithms to automatically analyze WSIs.
Convolutional Neural Networks (CNN) are the state-of-the-art algorithm for most of the computational pathology tasks, reaching high performance in tasks such as classification and segmentation.
Stain colour heterogeneity hinders the development of CNNs that generalize well on unseen data, especially when data originate from several medical centers~\cite{TLB2019,VFF2020,RHS2019,KAO2020,LPE2019}.
CNN training aims to learn relevant features for solving a task despite the stain variations between the images~\cite{HFK2020}.
However, CNNs usually underperform when tested on data with different colour variations than the ones used to train the network~\cite{TLB2019}, leading to models that are not effective when using data originating from several medical centers~\cite{RHS2019}.
This limitation increases the strain needed to develop models that produce consistent results and generalize well on heterogeneous data~\cite{KAO2020}, with significant effects on the algorithms reliability~\cite{OAA2019}.

Stain colour heterogeneity is still an unsolved problem in computational pathology, despite of the increasing number of methods developed to train CNNs that generalize better on stain heterogeneous data.
Methods to tackle stain variations involve mainly two approaches: approaches using using (1) the pixel-space and (2) the feature-space~\cite{VFF2020}.
Pixel-space methods include techniques to modify the characteristics of the input images~\cite{CGB2017,TLB2019,OAA2019}, such as colour normalization and colour augmentation.
Colour normalization applies a transformation to the image in order to match the stain of another image template~\cite{CGB2017}.
Traditional colour normalization approaches aim to estimate a colour matrix, that identifies the H\&E components of the image, such as~\cite{MNM2009,VPS2016,RAG2001}.
More recent approaches, such as~\cite{CLC2017,SBN2019,KLF2020,BeH2017}, involve Generative Adversarial Networks (GAN) to normalize the images, learning to match different stain distributions.
Colour augmentation applies a random perturbation to the image in order to have several stain variations during training, working on the matrix that describes H\&E components of the image or on the brightness and contrast of the image~\cite{TLB2019}.
Among these two techniques, colour augmentation has shown to often reach higher performance~\cite{TLB2019}.
However, the augmentation requires to tune the transformation parameters to avoid using artefacts to train the network, since the perturbation is often random.
Feature-space methods include techniques, such as domain-adversarial networks~\cite{OAA2019,RHS2019,HFK2020,LPE2019}, that force the CNN to learn domain-invariant features during the training.
Domain-adversarial networks work under the assumption that images coming from the same domain (such as the medical center, the patient or the tissue) share the same staining.
The CNNs, in parallel to the main task, learn features that are invariant to the domain where the image originates from~\cite{GUA2016} and as a consequence invariant to the stain variations.
Domain-adversarial networks have often shown higher performance when compared with colour augmentation, such as in~\cite{OAA2019,LPE2019}.
However, the definition of domain can be too strict, e.g. in the case where each patient is considered a domain~\cite{HFK2020}, or too broad, e.g. in the case where each medical center is considered a domain~\cite{OAA2019,LPE2019}.
In the first case, images from different patients can have the same stain; in the latter case, images coming from the same medical center can include different stains.

This paper aims to describe a novel integrated method to train CNNs that better generalize on data including images with colour variations that are different from the ones used for training. 
The method is based on a multi-task CNN, referred as H\&E-Adversarial Network (H\&E-AN).
H\&E-AN includes two output branches: the first one classifies histopathology images, while the second one predicts H\&E matrices linked to the images.
H\&E-AN training involves the optimization of a loss function including two terms, in order to learn discriminative and stain-invariant features through the adversarial optimization, similar to the one presented in domain-adversarial CNNs~\cite{GUA2016,OAA2019,LPE2019,HFK2020}. 
The adversarial optimization goes in the opposite gradient direction that minimizes the patch-classifier loss (to learn discriminative features respect to the classes) and in the positive direction of the gradient for the H\&E features, to maximize the H\&E-regressor loss.

The H\&E-AN CNN is evaluated on colon and prostate cancer WSI classification, comparing it with other algorithms developed to alleviate stain colour heterogeneity.
Colon cancer is the fourth most commonly diagnosed cancer in the world~\cite{BVA2018}.
Colon cancer diagnosis involves the detection of malignant glands and small agglomerations of cells~\cite{FMH2017}, called polyps, on the colon border.
Prostate cancer (PCa) is the second most frequent cancer in the male population worldwide~\cite{Raw2019}. 
Prostate cancer is diagnosed using the Gleason grading system~\cite{Bos1994} that involves the detection and the estimation of malignant glands in order to estimate the aggressiveness of cancer.
The comparison is made testing the methods on data originating from several unseen medical centers, to assess the CNN capability to generalize on images including stain variations.
H\&E-AN CNN reaches higher performance in classifying both prostate and colon images, compared with other methods to handle histopathology stain variability, showing that the method can help better deal with stain colour heterogeneous images.

\section{Methods}
\begin{figure*}[h]
\centering
\includegraphics[width=\linewidth]{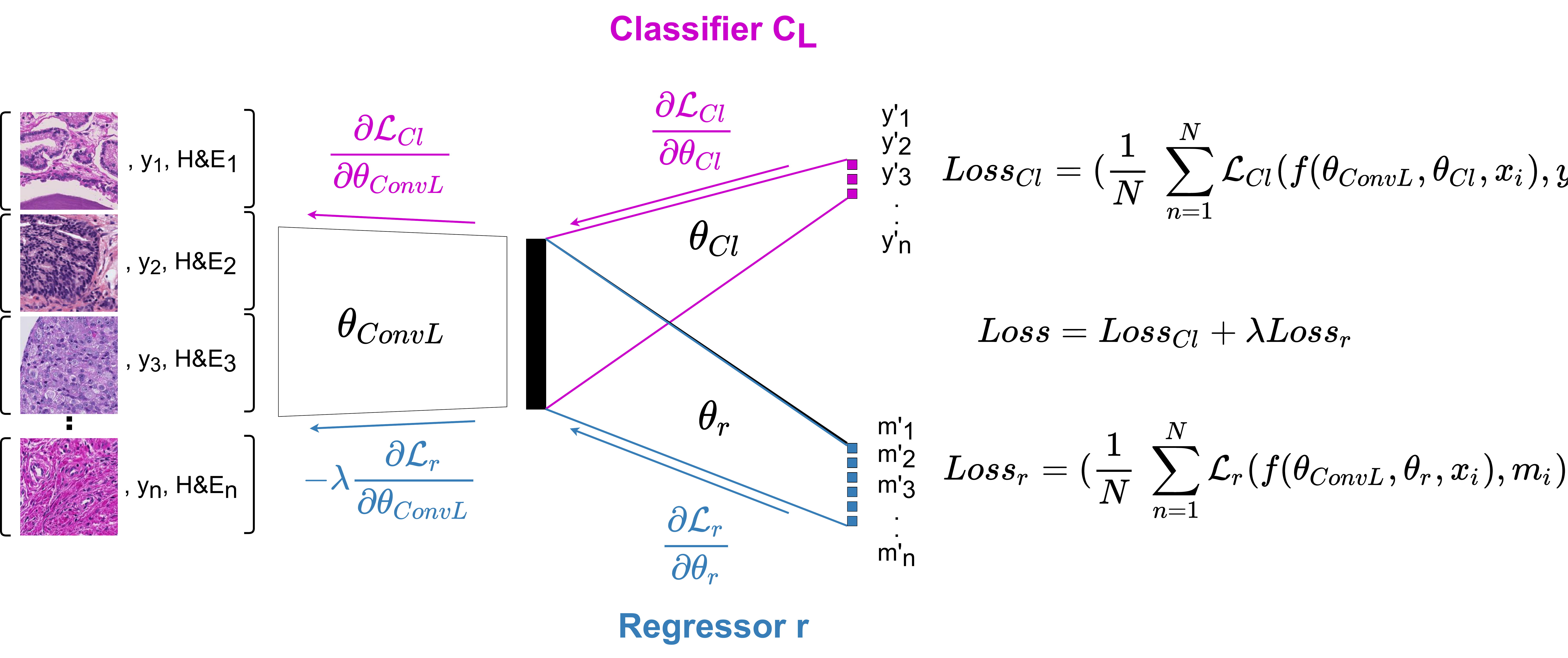}
\caption{Overview of the H\&E-adversarial CNN and of the training schema. The network includes three main components: convolutional layers $ConvL$, that produce discriminative and stain-invariant features, the classifier $Cl$ (purple) that predicts the class for the patches and the regressor $r$ (blue) that predicts the H\&E matrix component of the patches. The training involves the optimization of two loss functions: the minimization of $Loss_{Cl}$ for the classifier and the maximization of $Loss_{r}$ for the regressor. The latter term of the loss function is maximized through a gradient reverse layer, that applies a negative scalar to the gradients during the backpropagation. This schema aims to obtain discriminative (thanks to the classifier) and stain-invariant (thanks to the regressor) features.}
\label{fig:CNN_schema} 
\end{figure*}

The paper proposes H\&E-AN CNN method that exploits the information related to H\&E of the WSIs during the training of a CNN in WSI patch classification, to learn stain-invariant features.
The information related to the H\&E composition used to stain the image is a matrix $M$ including the RGB components of each reagent.

M =
$
\begin{pmatrix}
H_R & H_G & H_B\\
E_R & E_G & E_B\\
\end{pmatrix}
$

For each of the patches, the H\&E components are evaluated using the method proposed by Macenko et al.~\cite{MNM2009}, for each patch in the training set.
Each element of the matrix includes values ranging from 0 to 1.
The CNN is a multi-task network $f(\theta_{CNN},X)$ \textrightarrow $(Y, M)$ with trainable parameters $\theta_{CNN}$, that aims to predict the classes $Y$ and the H\&E components $M$ from the input patches $X$.
The CNN includes three main components: a convolutional layer block ($ConvL$), a classifier ($Cl$) and a regressor ($r$).
The convolutional layers are a function $f(\theta_{ConvL},x_{i})$ \textrightarrow $feat$ with trainable parameters $\theta_{ConvL}$ that produces a feature vector $feat_i$ from an input patch $x_i$. 
The classifier is a function $f(\theta_{Cl},feat_{i})$ \textrightarrow $y'_{i}$ with trainable parameters $\theta_{Cl}$ that makes class predictions $y'_{i}$ from $feat_i$. 
The regressor is a function $f(\theta_{r},feat_{i})$ \textrightarrow $m'_i$ with trainable parameters $\theta_{r}$ that predicts the H\&E matrix component $m'_{i}$ from $feat_i$. 

The training of the network aims to learn features that 1) reach high performance in patches classification and 2) do not take into account the stain variations of the training samples.
Therefore, the CNN must learn a feature representation $f(\theta_{ConvL})$ of the input data that is discriminative and stain-invariant~\cite{GaL2015,OAA2019}.
The training of the CNN involves the optimization of $\theta_{ConvL}$, $\theta_{Cl}$ and $\theta_{r}$.
For the training, the loss function $Loss$ involves two terms: $Loss_{Cl}$, that measures the error in classification task and $Loss_{r}$, that measures the error in the regression task.
The losses are described as follows (where $\mathcal{L}_{Cl}$ and $\mathcal{L}_{r}$ are a measure of discrepancy: 
\noindent 
\begin{eqnarray}
Loss_{Cl} = (\frac{1}{N}\  \sum\limits_{n=1}^N \mathcal{L}_{Cl}(f(\theta_{ConvL}, \theta_{Cl}, x_i), y_i)  \\
Loss_{r} = (\frac{1}{N}\  \sum\limits_{n=1}^N \mathcal{L}_{r}(f(\theta_{ConvL}, \theta_{r}, x_i), m_i)  \\
Loss = Loss_{Cl} + \lambda * Loss_r
\end{eqnarray}

The equations show that both the classifier and the regressor share the same set of features $\theta_{ConvL}$, that should be both discriminative with respect to the classes and stain-invariance.
The mechanism to have discriminative features is to minimize the $Loss_{Cl}$ (using Cross-Entropy as $Loss_{Cl}$) term of the loss function.
The mechanism to obtain stain-invariant features is to maximize the $Loss_{r}$ (using squared L2-Loss as $Loss_{r}$) of the loss function, as proposed in domain-adversarial networks~\cite{GUA2016,LPE2019,OAA2019,RHS2019,HFK2020}.
While the minimization of regression loss function would lead to learn the features to discriminate among stain variations, its maximization leads to learn stain-invariant features, so that the stain variations are indistinguishable for the model.
The squared L2-Loss function is chosen considering that the $M$ space does not include outliers that can hinder the regression, since the values in $M$ range from 0 to 1.
Regression term maximization is achieved by modifying the CNN, including a layer that reverses the gradients (through the multiplication of the gradient with a negative scalar $\lambda$) during the backpropagation and leave them unchanged during the forward propagation~\cite{GaL2015}.
The network iterative stochastic gradient mechanism for respectively $\theta_{ConvL}$, $\theta_{Cl}$, $\theta_{r}$ ($\mu$ is the learning rate) is described as follows:

\noindent 
\begin{eqnarray}
\theta_{ConvL} \leftarrow \theta_{ConvL} - \mu (\frac{\partial \mathcal{L}_{Cl}^i}{\partial \theta_{ConvL}} - \lambda \frac{\partial \mathcal{L}_r^i}{\partial \theta_{ConvL}}) \\
\theta_{Cl} \leftarrow \theta_{Cl} - \mu (\frac{\partial \mathcal{L}_{Cl}^i}{\partial \theta_{Cl}}) \\
\theta_{r} \leftarrow \theta_{r} - \lambda \mu (\frac{\partial \mathcal{L}_{r}^i}{\partial \theta_{r}}) 
\end{eqnarray}

The CNN architecture is shown in Figure~\ref{fig:CNN_schema}.

\begin{table}[h]
\caption{Composition of the colon dataset. The colon dataset originates from seven medical sources and it is annotated with three classes: Cancer, Dysplasia and Normal. Data originating from hospitals (AOEC and Radbouducm) are used to train, validate and as internal test set for the CNN. Data originating from publicly available repositories are used as external test set to evaluate the capability of the CNN to generalize on stain heterogeneous data.}
\resizebox{0.5\textwidth}{!}{
\begin{tabular}{|lllll|}
\hline
\textbf{Medical Source}            & \textbf{Cancer} & \textbf{Dysplasia} & \textbf{Normal} & \textbf{Total} \\ \hline
\textbf{AOEC}     & \textbf{6158}   & \textbf{23312}     & \textbf{3853}   &       \textbf{33323}         \\ \hline
Training                       & 4059            & 13170              & 3402            &        20631        \\
Validation                     & 844             & 4005               & 78              &         4927       \\
Testing                        & 1255            & 6137               & 373             &        7765        \\ \hline
\textbf{Radboudumc}     & \textbf{4430}   & \textbf{3542}      & \textbf{1998}   &    \textbf{9970}           \\ \hline
Training                       & 2995            & 2498               & 1304            &        6797        \\
Validation                     & 643             & 707                & 365             &        1715        \\
Testing                        & 792             & 337                & 329             &         1458       \\ \hline
\textbf{Internal Testing data} & \textbf{2047}   & \textbf{6474}      & \textbf{702}    &       \textbf{9970}         \\ \hline
AIDA                           & 7881            & 3296               & 31859           &       43036         \\
GlaS                           & 450             & 0                  & 210             &          660      \\
CRC                            & 1507            & 0                  & 1144            &         2651       \\
UNITOPATHO                          & 0               & 18064              & 3487            &         21551       \\
CAMEL                          & 0               & 15757              & 12030           &       27787         \\ \hline
\textbf{External Testing data} & \textbf{9838}   & \textbf{37117}     & \textbf{48730}  &      \textbf{95685}          \\ \hline
\end{tabular}
}
\label{tbl:colon_data}
\end{table}

\begin{table}[h]
\caption{Composition of the prostate dataset. The prostate dataset originates from four medical sources and it is annotated with four classes: Benign, Gleason Pattern 3 (GP3), Gleason pattern 4 (GP4) and Gleason pattern 5 (GP5). The TMAZ and Sicapv2 datasets are used to train, validate and as internal test set for the CNN. Gleason challenge and DiagSet datasets are used as external test set, to evaluate the capability of the CNN to generalize on stain heterogeneous data.}
\resizebox{0.5\textwidth}{!}{
\begin{tabular}{|cccccc|}
\hline
\textbf{Medical Source}            & \textbf{Benign} & \textbf{GP3} & \textbf{GP4} & \textbf{GP5} & \textbf{Total} \\ \hline
\textbf{TMAZ}                  & \textbf{3487}   & \textbf{8946}              & \textbf{7424}              & \textbf{3610}              & \textbf{23467} \\ \hline
Training                       & 2010            & 5992                       & 4472                       & 2766                       & 15240          \\
Validation                     & 1350            & 1352                       & 831                        & 457                        & 4927           \\
Testing                        & 127             & 1602                       & 2121                       & 387                        & 4237           \\ \hline
\textbf{Sicapv2}               & \textbf{11069}  & \textbf{10784}             & \textbf{2979}              & \textbf{2767}              & \textbf{27599} \\ \hline
Training                       & 9432            & 6499                       & 2250                       & 2011                       & 20192          \\
Validation                     & 604             & 819                        & 302                        & 210                        & 1935           \\
Testing                        & 1033            & 3466                       & 427                        & 546                        & 5472           \\ \hline
\textbf{Internal Testing Data} & \textbf{1160}   & \textbf{5068}              & \textbf{2548}              & \textbf{933}               & \textbf{9709}  \\ \hline
Gleason Challenge              & 1080            & 2431                       & 3649                       & 100                        & 7260           \\
Diagset                        & 8783            & 1243                       & 4334                       & 696                        & 15056          \\ \hline
\textbf{External Testing Data} & \textbf{9863}   & \textbf{3674}              & \textbf{7983}              & \textbf{796}               & \textbf{22316} \\ \hline
\end{tabular}
}
\label{tbl:prostate_data}
\end{table}

\begin{figure}[h!]
\centering
\includegraphics[width=\linewidth]{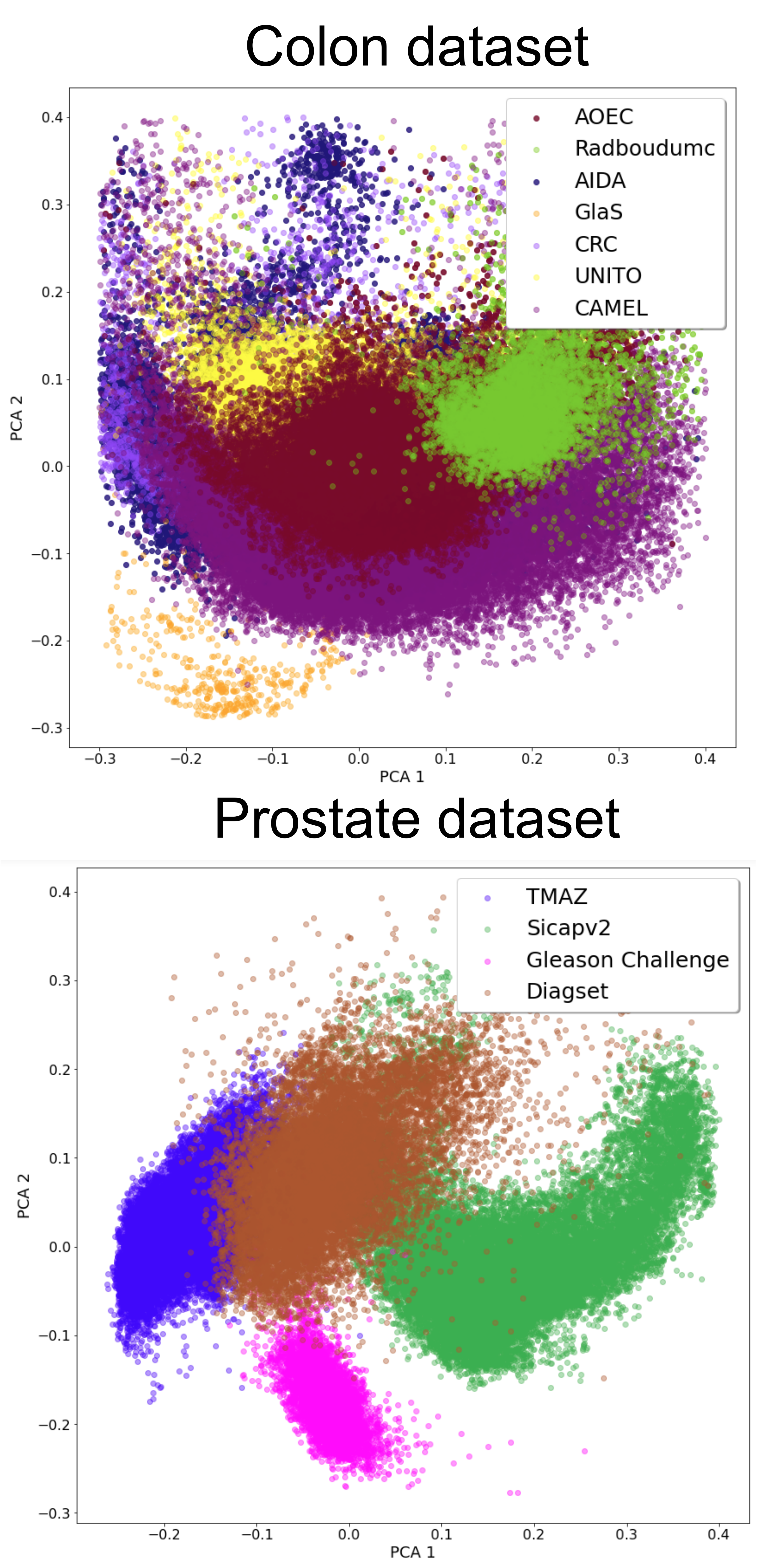}
\caption{Stain heterogeneity across datasets. The Figure shows the H\&E components of the patches for colon (upper part) and the prostate (lower part) datasets. The components are projected in a bidimensional space through PCA analysis. Each dot is a patch, highlighted with a different colour according to the source dataset. }
\label{fig:heterogeneity_distribution} 
\end{figure}
\section{Experiments}
\paragraph{Datasets}
The method is trained and evaluated on highly-heterogeneous histopathological images, including colon and prostate tissue.
In both partitions, the images originate from several medical sources, guaranteeing high heterogeneity in terms of stain and allowing to evaluate the capability of the CNN to generalize on images with unseen stain variations.
Image heterogeneity is shown in Figure~\ref{fig:heterogeneity_images} and Figure~\ref{fig:heterogeneity_distribution}.
As a basic overview, Figure~\ref{fig:heterogeneity_distribution} shows the distribution of the patches according to their H\&E components. The components, highlighted with a different colour to identify the dataset where they originate from, are projected on a bidimensional space using the Principal Component Analysis (PCA)~\cite{Jol2005}.

The colon partition includes images collected from seven medical sources.
The images are WSIs (private dataset from Azienda Ospedaliera per le Emergenze e Urgenze Cannizzaro, AOEC, private dataset from Radboudumc, Analytic Imaging Diagnostics Arena~\cite{SLL2021}, AIDA) and small cropped sections of WSIs (Gland Segmentation in Colon Histology Images Challenge~\cite{SSR2015}, GlaS, ColoRectal Cancer Tissue Phenotyping~\cite{ASE2017}, CRC, UNITOPATHO~\cite{BPT2021,BBP2021}, CAMEL dataset~\cite{XSS2019}).
The WSIs are pixel-wise annotated by pathologists, while the cropped sections are annotated at section level.
Among all the annotations available, the classes chosen to test the method are cancer, dysplasia and normal tissue.
AOEC and Radboudumc private datasets originate from two hospitals and are used to train, validate and test the method, as internal testing data.
The other five datasets originate from publicly available repositories and are used as external test data.
Table~\ref{tbl:colon_data} summarizes the colon partition composition.

The prostate partition includes images collected from four publicly available datasets.
The images are Tissue Micro Arrays (TMAs) (The Tissue Micro Aarray Zurich~\cite{AFM2018}, TMAZ, and Gleason Challenge~\cite{NHK2018,KNF2019}) and WSIs (Sicapv2~\cite{SCS2020} and DiagSet~\cite{KCO2021}).
The images are pixel-wise annotated by pathologists, with benign, Gleason Pattern 3 (GP3), Gleason Pattern 4 (GP4) and Gleason Pattern 5 as classes.
TMAZ and Sicapv2 and are used to train, validate and test the method, as internal testing data.
Gleason Challenge and DiagSet are are used as external test data.
Table~\ref{tbl:prostate_data} summarizes the prostate partition composition.
\paragraph{Pre-processing}
The image-preprocessing involves WSI splitting into patches and the extraction of the patches coming from tissue regions, for both colon and prostate.
The patches are 224x224 pixels in size, in order to facilitate the input for the CNN.
The colon dataset includes WSIs and cropped sections of WSIs.
The WSIs come with a tissue mask, including pixel-wise annotations. 
WSIs are split into a grid of patches at magnification 10x.
The grid is built based on the highest magnification level available $M_m$, with patches of $p_s$ in size.
Afterwards, grid patches are resized to 224x224, using the following equation, as shown in~\cite{MOP2021}:
\begin{eqnarray}
224 : 10 = p_s : M_m
\end{eqnarray}
The cropped sections come without a tissue mask, but being small images, the section's label is assigned to the tissue.
In order to avoid the extraction of patches from background regions, a tissue mask is generated using the method proposed by~\cite{RHA2018}.
Patches are randomly extracted, with a variable number depending on the size of the images, to avoid strong overlap of the patches: 20 from CRC, 2 from GlaS, 5 from UNITO and CAMEL. 
The numbers are chosen considering the amount of pixels included in each of the datasets.

The prostate dataset includes pixel-wise annotated TMAs and WSIs.
The patches are extracted with a size of 750x750 pixels from magnification 40x and resized to 224x224, as shown in~\cite{AFM2018,NOM2021}.
In TMAs (TMAZ and Gleason Challenge), 30 patches are randomly extracted from each core.
In WSIs (Sicavp2 and Diagset), the images are split into a grid of patches from the highest magnification level available, as shown for colon WSIs, using the following equation:
\begin{eqnarray}
750 : 40 = p_s : M_m
\end{eqnarray}
%
\paragraph{Comparison with other methods}
The method proposed in this paper is compared with other methods developed to train CNNs that generalize on data with high stain heterogeneity.
All the methods are evaluated using quadratic weighted Cohen's Kappa score ($\kappa$-score)~\cite{McH2012} as metric and the Wilcoxon Rank-Sum test~\cite{Wil1992} is applied to check if the improvement obtained by the H\&E-AN is statistically significant.
The comparison involves six methods: a CNN trained without any augmentation or normalization of the images, a CNN trained using a stain normalization method, a CNN trained using stain normalization through a StainGAN, a CNN trained using colour augmentation methods, a CNN trained using stain augmentation, a CNN trained using a domain-adversarial network.
For each method, the performance is evaluated on the internal test set, on the external test and on their combination.
The stain normalization method is the one proposed by Macenko et al.~\cite{MNM2009}.
For each type of tissue, a target image is randomly selected so that the stains of all the images in training and testing sets matches the target image's stain.
The StainGAN network is the one proposed by Shaban et al.~\cite{SBN2019}, to normalize images from different domains.
In this case, the StainGAN is pre-trained to normalize images through the training datasets, for colon and prostate.
Afterwards, the network is used to normalize images during CNN training.
The stain augmentation method involves the perturbation of the colours within the images~\cite{TLB2019}.
The parameters involved in the random perturbation are the brightness, hue, saturation.
The stain augmentation method involves the perturbation of the H\&E matrix~\cite{TLB2019}.
Two parameters are involved: a random constant $\sigma_1$ (multiplied to the H\&E matrix) that scales the colour intensity and a random constant $\sigma_2$ (added to the H\&E matrix * $\sigma_1$ product) that increase or reduce the colour intensity.
The domain-adversarial CNN is the one proposed by Otalora et al.~\cite{OAA2019}.
The network is a multi-task CNN, that predicts the patches and predicts the domain where they originate from.
As described for the StainGAN implementation, the domain-adversarial CNN is trained using data from two domains.

The $\kappa$-score is a metric that measures the reliability between annotations, commonly used in histopathology to assess the pathologists' performance in the evaluation of images.
In this case, the annotations are the predictions made by the CNN and the ground truth made by pathologists.
The optimal value of the $\kappa$-score is 1, which means a complete agreement between the annotations, while a $\kappa$-score equal -1 means a complete disagreement between the annotations.
$\kappa$-score equal 0 means random agreement, as the measure is normalized with respect to agreement by chance.

The Wilcoxon Rank-Sum test is used to assess if two probabilistic populations have the same distribution (null hypothesis).
The null hypothesis is tested positive when the $p$-value $>$ 0.05, while it is tested negative (or rejected) when the $p$-value $<$ 0.05.
In this case, the comparison is made between the H\&E-AN and the method that reaches the highest performance among the six other methods evaluated. 
If the null hypothesis is tested negative, the improvement is statistically significant.
\begin{table*}[]
\caption{Overview of the CNN results on the colon test partitions, assessed with the $\kappa$-score for each of the methods tested. The results that are statistically significant (compared with the method reaching the highest performance among the ones used as comparison) are reported with an asterisk (*).}
\centering
\begin{tabular}{|l|l|l|l|}
\hline
\textbf{Method}               & \textbf{Internal Test Set} & \textbf{External Test Set} & \textbf{Test Set} \\ \hline
no augmentation               &   0.644 $\pm$ 0.037                         &   0.424 $\pm$ 0.074                         &    0.453 $\pm$ 0.075               \\ \hline
stain normalization~\cite{MNM2009} &   0.650 $\pm$ 0.036                         &   0.492 $\pm$ 0.037                         &     0.510 $\pm$ 0.034              \\ \hline
StainGAN~\cite{SBN2019}             &   0.474 $\pm$ 0.062                         &        0.463 $\pm$ 0.053                    &    0.487 $\pm$ 0.051               \\ \hline
colour augmentation            &    0.636 $\pm$ 0.032                        &          0.488 $\pm$ 0.049                  &       0.518 $\pm$ 0.046            \\ \hline
stain augmentation            &    0.655 $\pm$ 0.033                        &          0.498 $\pm$ 0.041                  &       0.515 $\pm$ 0.034            \\ \hline
domain adversarial~\cite{OAA2019}            &     0.621 $\pm$ 0.044                       &    0.497 $\pm$ 0.039                        &     0.513 $\pm$ 0.046              \\ \hline
\textbf{Our Method}                    &       \textbf{0.661 $\pm$ 0.030}                     &    \textbf{0.532 $\pm$ 0.038*}                        &        \textbf{0.556 $\pm$ 0.040*}           \\ \hline
\end{tabular}
\label{tbl:colon_results}
\end{table*}

\begin{table*}[]
\caption{Overview of the CNN results on the prostate test partitions, assessed with the $\kappa$-score for each of the methods tested. The results that are statistically significant (compared with the method reaching highest performance among the ones used as comparison) are reported with an asterisk (*).}
\centering
\begin{tabular}{|l|l|l|l|}
\hline
\textbf{Method}               & \textbf{Internal Test Set} & \textbf{External Test Set} & \textbf{Test Set} \\ \hline
no augmentation               &   0.712 $\pm$ 0.026                         &   0.052 $\pm$ 0.099                         &    0.142 $\pm$ 0.081               \\ \hline
stain normalization~\cite{MNM2009} &   0.679 $\pm$ 0.066                         &   0.271 $\pm$ 0.101                         &     0.336 $\pm$ 0.078              \\ \hline
StainGAN~\cite{SBN2019}             &   0.633 $\pm$ 0.065                         &        0.244 $\pm$ 0.113                   &    0.344 $\pm$ 0.080              \\ \hline
colour augmentation            &    0.714 $\pm$ 0.044                        &          0.418 $\pm$ 0.043                  &       0.476 $\pm$ 0.040            \\ \hline
stain augmentation            &    0.687 $\pm$ 0.040                        &          0.312 $\pm$ 0.104                  &       0.298 $\pm$ 0.086            \\ \hline
domain adversarial~\cite{OAA2019}            &     0.670 $\pm$ 0.071                       &    0.390 $\pm$ 0.110                        &     0.449 $\pm$ 0.098              \\ \hline
\textbf{Our Method}                    &       \textbf{0.725 $\pm$ 0.035}                     &    \textbf{0.474 $\pm$ 0.066*}                         &        \textbf{0.532 $\pm$ 0.057*}           \\ \hline
\end{tabular}
\label{tbl:prostate_results}
\end{table*}

\paragraph{Experimental parameters}
The proposed H\&E-adversarial CNN and the methods used for comparison are implemented on the same backbone architecture and are trained multiple times, adopting the same strategy to set the hyper-parameters and to face the class imbalance.
The backbone architecture is a DenseNet121 (pre-trained on ImageNet).
The network produces a feature vector of size 1024 for each input patch. An intermediate fully connected layer with 128 nodes is inserted between the classifier and the output of the classifier.
For each method, the CNN is trained ten times (average and standard deviation of the models are reported) to limit the non-deterministic effect of the stochastic gradient descent used to optimize the model.
CNN hyper-parameters are chosen after the evaluation of a grid search algorithm~\cite{Chi2017}.
Grid search aims to find an optimal configuration for the CNN hyperparameters.
In this case, the optimal configuration is the one that allows the CNN to have the lowest loss function on the validation partition.
The hyper-parameters involved in the grid search are the optimizer (Adam), the learning rate ($10^{-3}$), the decay rate ($10^{-4}$), the number of epochs (15 epochs with an early stop mechanism used to stop the training if the validation loss function has not decreased for more than five epochs), the number of nodes within the intermediate fully-connected layers (128), $\sigma_1$ and $\sigma_2$ for the stain augmentation algorithm (both 0.2), the $\lambda$ parameter for the domain-adversarial CNN and for H\&E-AN (respectively 0.5 and 1).
Moreover, the grid search algorithm is adopted to identify the shifts for colour algorithm: in the colon data the hue shift is limited to be between -15 and 8, the saturation shift between -20 and 10 and the brightness shift between -8 and 8, while in prostate data the hue shift is limited to be between -9 and 9, the saturation shift is limited to be between -25 and 25 and the brightness shift is limited to be between -10 and 10.
The effects of class imbalance are handled adopting a class-wise data augmentation method, involving three operations: rotations, flipping and colour augmentation.
Data augmentation is implemented with the Albumentations library~\cite{BPK2018}.
\begin{figure*}[ht]
\centering
\includegraphics[width=0.9\linewidth]{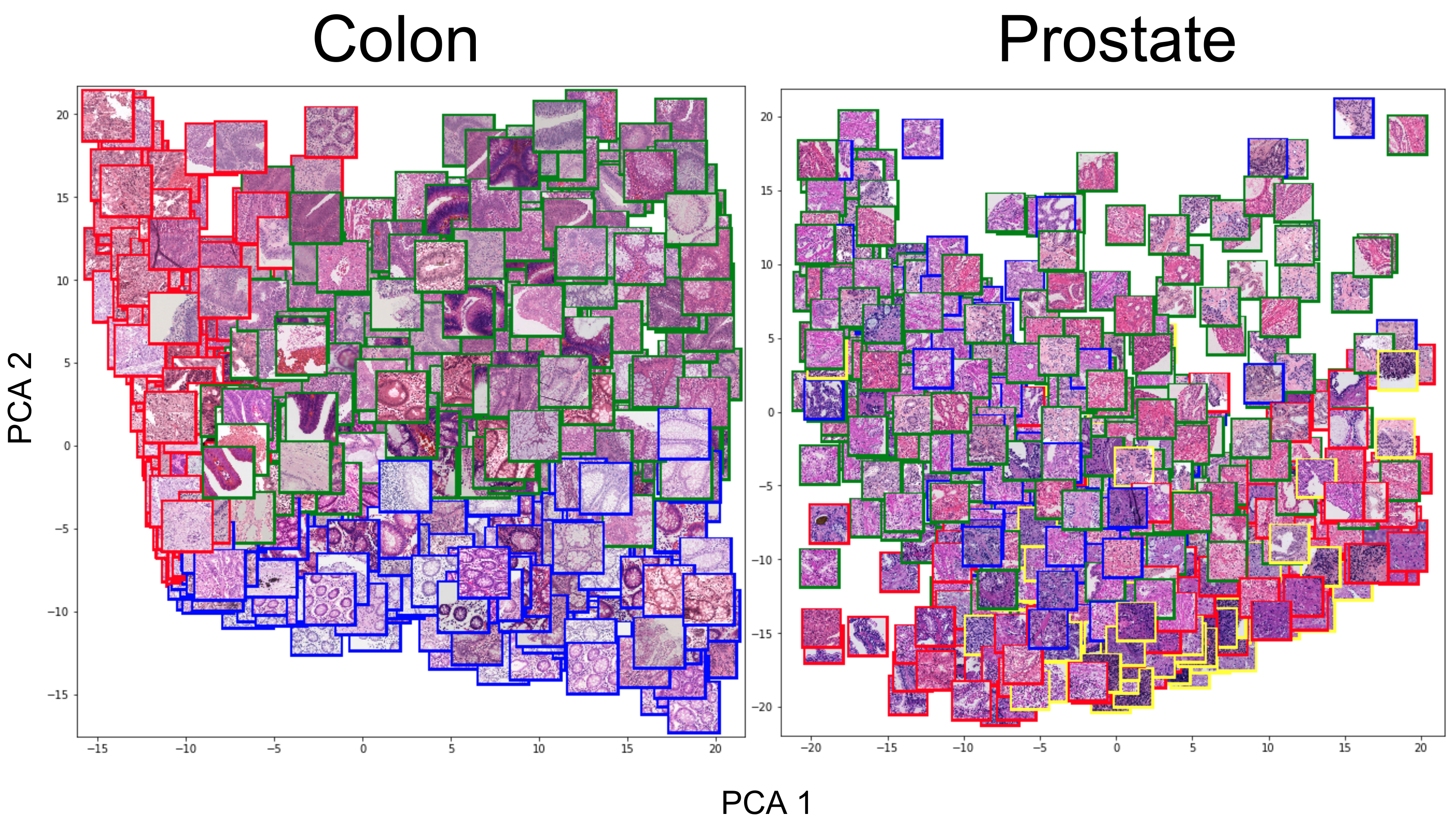}
\caption{Stain-invariant feature representation of the images. The figure shows the distribution learnt by the CNN for the colon patches (left) and prostate (right), projecting the features in a bidimensional space through the PCA algorithm. The features are invariant to the stain, since it is possible to identify regions including a structure, such as glands, where several stain variations are present. The patches are highlighted with colour corresponding to the predicted class. In the colon subfigure: red is cancer, green is dysplasia, blue means normal glands. In prostate: red is benign, green is Gleason pattern 3, blue is Gleason pattern 4, yellow is Gleason pattern 5.}
\label{fig:feature_representation} 
\end{figure*}

\section{Results}
The H\&E-AN outperforms the CNNs trained using other methods, demonstrating the capability to better alleviate stain heterogeneity in colon and prostate histopathology image classification.

The evaluation on colon data involves the classification of cancer, dysplasia and normal glands on the internal test set (including AOEC and Radboudumc test partitions), the external test set (including AIDA, GlaS, CRC, UNITO and CAMEL datasets) and the cumulative test set, including the combination of the internal and external test sets.
The H\&E-AN outperforms the other methods in each of the colon test sets.
While in the internal test set the performance of the method is comparable with the ones of the other methods, on both the external and cumulative test sets the H\&E-AN method obtains statistically significant improvements over the other methods. The CNN reaches respectively a $\kappa$-score = $0.532\pm0.038$ (higher than the best among the other methods used for the comparison, the CNN trained with stain augmentation on the external test set) and $\kappa$-score = $0.556\pm0.040$ (higher than the best among the other methods used for the comparison, the CNN trained with stain augmentation on the cumulative test set).

The evaluation on prostate involves the classification of benign, Gleason pattern 3 (GP3), Gleason pattern 4 (GP4) and Gleason pattern 5 (GP5) on the internal test set (including the TMAZ and Sicapv2 test partitions), the external test set (including Gleason Challenge and DiagSet datasets) and the cumulative test set, including the combination of the internal and external test sets.
The H\&E-AN outperforms the other methods in each of the prostate test sets.
While on the internal test set the performance of the method is comparable with the ones of the other methods, in both the external and cumulative test sets H\&E-AN the method obtains statistically significant improvements over the other methods.
The CNN reaches respectively a $\kappa$-score = $0.474\pm0.066$ (higher than the best among the other methods used for the comparison, colour augmentation, on the external test set) and $\kappa$-score = $0.532\pm0.057$ (higher than the best among the other methods used for the comparison, colour augmentation, on the cumulative test set).

\section{Discussion}
The results show that the H\&E-AN obtains higher performance compared with the other methods proposed in scientific literature to train CNNs that generalize on stain heterogeneous data, learning a stain-invariant feature representation.
The H\&E-AN reaches the highest performance in both colon and prostate image classification, considering both the internal and the external test set.
The result suggests that the model generalizes better on datasets including heterogeneous stain variations.
This fact can be explained considering the training mechanism of the multi-task network.
During the training, the loss functions measuring the error in classification is minimized, while the loss function measuring the error in the regression of the H\&E matrix is maximized.
This loss function encourages the CNN to learn features that are determinant for the classification (through the minimization of classifier loss) and that are not dependent on the stain variations (through maximization of the regressor).
Domain-adversarial CNNs inspire the mechanism.
However, domain-adversarial CNNs rely on the domain concept, which may be fuzzy considering that images from the same domain may have different stain variations or that images from different domains may share the same stain variations.
On the other hand, the regression of the H\&E matrix allows the model to directly learn features invariant to the stain instead of features invariant to the domain.
The internal test set (including data coming from the same repository used to train the CNNs) and the external one (including data coming from unseen repositories) are heterogeneous.
This fact can be seen looking at Figure~\ref{fig:heterogeneity_distribution}, which shows the patch distribution highlighting the dataset source and considering the results reached by the CNN trained without any normalization or augmentation.
Without keeping into account data heterogeneity, the CNN reaches good performance on the internal test set, where the stain variations are similar to the ones within the training set.
However, the CNN underperforms (on prostate data the model can be considered as a random classifier) when tested on the external test set, including different stain variations.

The fact that the H\&E-AN reaches the highest performance on the external data can be explained with the fact that the CNN learns stain-invariant features during the training, as shown in Figure~\ref{fig:feature_representation}.
The Figure shows the feature representation of the H\&E-AN, for both colon and prostate images.
The features (from $ConvL$ block) of each patch are projected on a bidimensional space, using the PCA algorithm.
Within the distribution, it is possible to identify regions including similar structures, such as healthy glands in colon (bottom right) or malignant glands infiltrated with cells in prostate (top left).
Those regions includes patches with different stain variations, that are positioned in the same zone.
\section{Conclusions}
This paper introduces a novel H\&E-AN to train CNNs that better generalize on stain heterogeneous data.
The method exploits the H\&E information of the images to learn stain invariant-features.
The method involves a multi-task CNN, including a classifier (to predict patch classes) and a regressor (to predict the H\&E matrix).
The H\&E-AN outperforms other methods considered in the comparison, including colour normalization, StainGAN, colour augmentation and domain-adversarial CNN, demonstrating its usefulness to handle stain variability in digital pathology images.
We plan to test H\&E-AN on other tasks (such as segmentation) and other datasets, evaluating the possibility to combine the method with other techniques and expanding the comparison to other methods.
The code with the H\&E-AN implementation is publicly available on Github (https://github.com/ilmaro8/HE\_adversarial\_CNN). 

{\small
\bibliographystyle{ieee_fullname}
\bibliography{egbib}
}

\end{document}